\documentclass[aps,prb,twocolumn,showpacs,superscriptaddress,longbibliography]{revtex4-1}

\usepackage{amssymb}
\usepackage{amsfonts}
\usepackage{amsmath}
\usepackage{bm}
\usepackage{textcomp}
\usepackage{color}
\usepackage{longtable}
\usepackage{graphicx}
\usepackage{dcolumn}
\usepackage[a4paper=true,pagebackref=false]{hyperref}

\begin{document}

\title{Nanosecond thermometry with Josephson junction}

\author{M.~Zgirski}
\affiliation{Institute of Physics, Polish Academy of Sciences, Aleja Lotnikow 32/46, PL 02668 Warsaw, Poland}

\author{M.~Foltyn}
\affiliation{Institute of Physics, Polish Academy of Sciences, Aleja Lotnikow 32/46, PL 02668 Warsaw, Poland}

\author{A.~Savin} \affiliation{Low Temperature Laboratory, Department of Applied Physics, Aalto University School of Science, P.O. Box 13500, 00076 Aalto, Finland}

\author{M.~Meschke} \affiliation{Low Temperature Laboratory, Department of Applied Physics, Aalto University School of Science, P.O. Box 13500, 00076 Aalto, Finland}

\author{J.~Pekola} \affiliation{Low Temperature Laboratory, Department of Applied Physics, Aalto University School of Science, P.O. Box 13500, 00076 Aalto, Finland}

\date{\today}

\begin{abstract}
We demonstrate a novel approach to thermometry at the nanoscale exploiting a superconducting weak link. Such a weak link probed with nanosecond current pulses serves as a temperature sensing element and, due to the fast inherent dynamics, is capable of delivering unprecedented temporal resolution. We employ the thermometer to measure dynamic temperature of electrons in a long superconducting wire relaxing to the bath temperature after application of the heating pulse. Our measurement delivers nanosecond resolution thus providing the proof-of-concept of the fastest-to-date all-solid-state thermometry. Our method improves the state-of-the-art temporal resolution of mesoscopic thermometry by at least two orders of magnitude, extending temporal resolution of existing experiments and introducing new possibilities for ultra-sensitive calorimeters and radiation detectors.
\end{abstract}
\maketitle

Investigations of thermal processes in mesoscopic systems demand application of a fast thermometry that can be easily integrated with a structure\cite{Gasparinetti2015,Cleland2004,Pekola2016}. With reduction of the volume of a thermodynamic system its thermal inertia rapidly vanishes leaving often very short time interval for observation of a transient from which all important thermodynamical quantities can be derived. Static methods employed normal metal-insulator-superconductor (N-I-S) tunnel junctions\cite{Feshchenko2015}, SQUID noise thermometry\cite{Beyer2007} or quantum dots\cite{Gasparinetti2011} to explore hot electron effects\cite{Clarke1994}, quantization of heat conductance\cite{Schwab2000,Pekola2006,Jezouin2013}, build Maxwell`s demons\cite{Koski2015} and microcoolers\cite{Nahum1994}. Some dynamical thermal properties were measured in steady states e.g. relaxation time of an excess electron energy to a phonon bath $\tau_{e-p}$ can be accessed by measuring electron-phonon thermal conductance $G_{e-p}$ and assuming the usual linear temperature dependence for the electronic heat capacity\cite{Clarke1994}. However to get a complete understanding of thermodynamics at nanoscale one needs obviously to have a thermometer operating at time scales much shorter than thermal relaxation times\cite{Cleland2003,Gasparinetti2015,Pekola2016}. Fast thermometry is prerequisite for time-resolved bolometers - detectors of electromagnetic radiation, especially in the far-infrared and THz band, for health\cite{Fitzgerald2006}, security\cite{Appleby2007} and astronomical applications\cite{Ferguson2002}. A typical thermal relaxation time $\tau$ of a nanoisland is the ratio of its heat capacity $C$ to thermal conductance $G$ providing a path to thermal reservoir for an excess energy. If such an island is used as a sensing element of a bolometer (e.g. absorbing single photons), to increase device sensitivity, it is highly desirable to reduce its heat capacity to maximize temperature rise upon photon absorption ($\Delta T=h\nu /C$) and reduce thermal conductance to reservoir. However, such an optimization may lead to reduction of the relaxation time of the nanoisland calling for the application of even faster thermometers. Fast thermometry would also lend strong support to development of cryoelectronics and quantum computing devices making it possible to control temperature of different components of the devices and monitor their thermal coupling to environment.

One approach to boost the temporal resolution of a thermometer is to embed a temperature sensor into a microwave or RF resonator\cite{Gasparinetti2015,Pekola2016,Cleland2003}. A change in magnitude and phase of transmitted or reflected signal bears information about thermal dynamics of the system. The method circumvents the problem of unavoidable stray cabling capacitance offering a typical bandwidth of 10\,MHz. The need to use a resonator increases the sensor complexity and inhibits a higher level of integration (microwave on-chip resonators are mm-sized structures). In an effort to explore thermal processes at significantly faster rates we have developed a completely different strategy: we employ a hysteretic superconducting weak link probed with fast current pulses for its switching threshold as a temperature sensing element. Our thermometer is capable of measuring temperature transients with unprecedented temporal resolution, being inherently limited only by plasma frequency of the Josephson junction (JJ), with response in the picoseconds range. Moreover, it also exhibits other valuable characteristics: (i) it is, to the best of our knowledge, the smallest all-solid-state-based thermometer; (ii) it is very simple to fabricate e.g. the Dayem nanobridge is just a piece of a nanowire interrupting a thicker wire; (iii) it can be easily integrated with different nanostructures providing high spatial resolution for the temperature read-out; and (iv) it requires much simpler hardware configuration compared to existing RF-techniques. The ease of integration, true nanometer size and simplicity make our thermometer a candidate for ultra-low energy calorimetry and bolometry applications. Switching thermometry, as described below, can prove to be very attractive in many physical experiments e.g. in determination of heat capacity, thermal conductivity, studying mechanisms of heat exchange in nanostructures or even in experiments detecting single photons, provided possibility to launch them on demand synchronized to the pulses probing the JJ.

Below we describe our approach to fast thermometry at nanoscale. First we show how the switching feature of any superconducting weak link i.e. its transition from the superconducting to the normal state can be utilized to derive the weak link temperature. We validate the introduced probing protocol by studying dynamic temperature of our model system (Aluminum superconducting nanowire), with true nanosecond resolution, and compare our measurement with prediction of the heat flow equation. Subsequently, prior to a summary, we outline the powerful perspectives for future studies that our method brings about.

\begin{figure}
\centering
\includegraphics[width=0.5\textwidth]{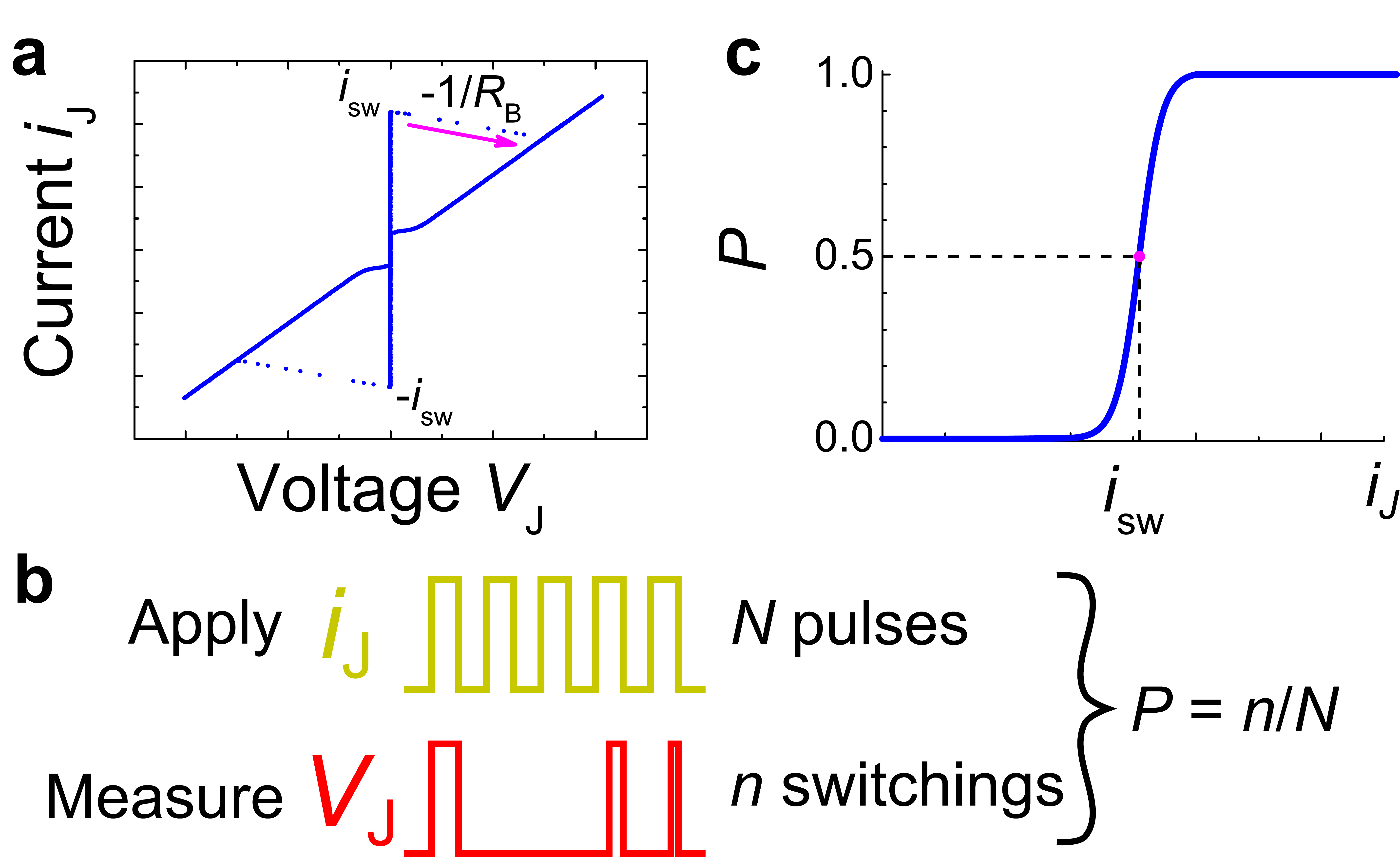}
\caption{\label{fig:Pulses}\textbf{Switching current measurement.} \textbf{(a)} IV characteristics of a JJ biased through $R_B$ bias resistor (cf. Fig.\,\ref{fig:Bridge}). The JJ supports supercurrent only to a certain level. On crossing the threshold value $i_{sw}$ a finite voltage develops across the JJ. Dots, revealing bias line with slope $-1/R_B$, are measuring artifacts related to finite response of room temperature electronics. In reality switching process is instantaneous. \textbf{(b)} An estimator for switching probability $P$ at given current amplitude $i_J$ is measured with a train of $N$ pulses. \textbf{(c)} An S-curve: $P(i_J)$ dependence.}
\end{figure}

\subsection*{Josephson junction as a temperature-sensitive switch}
\begin{figure}
\centering
\includegraphics[width=0.5\textwidth]{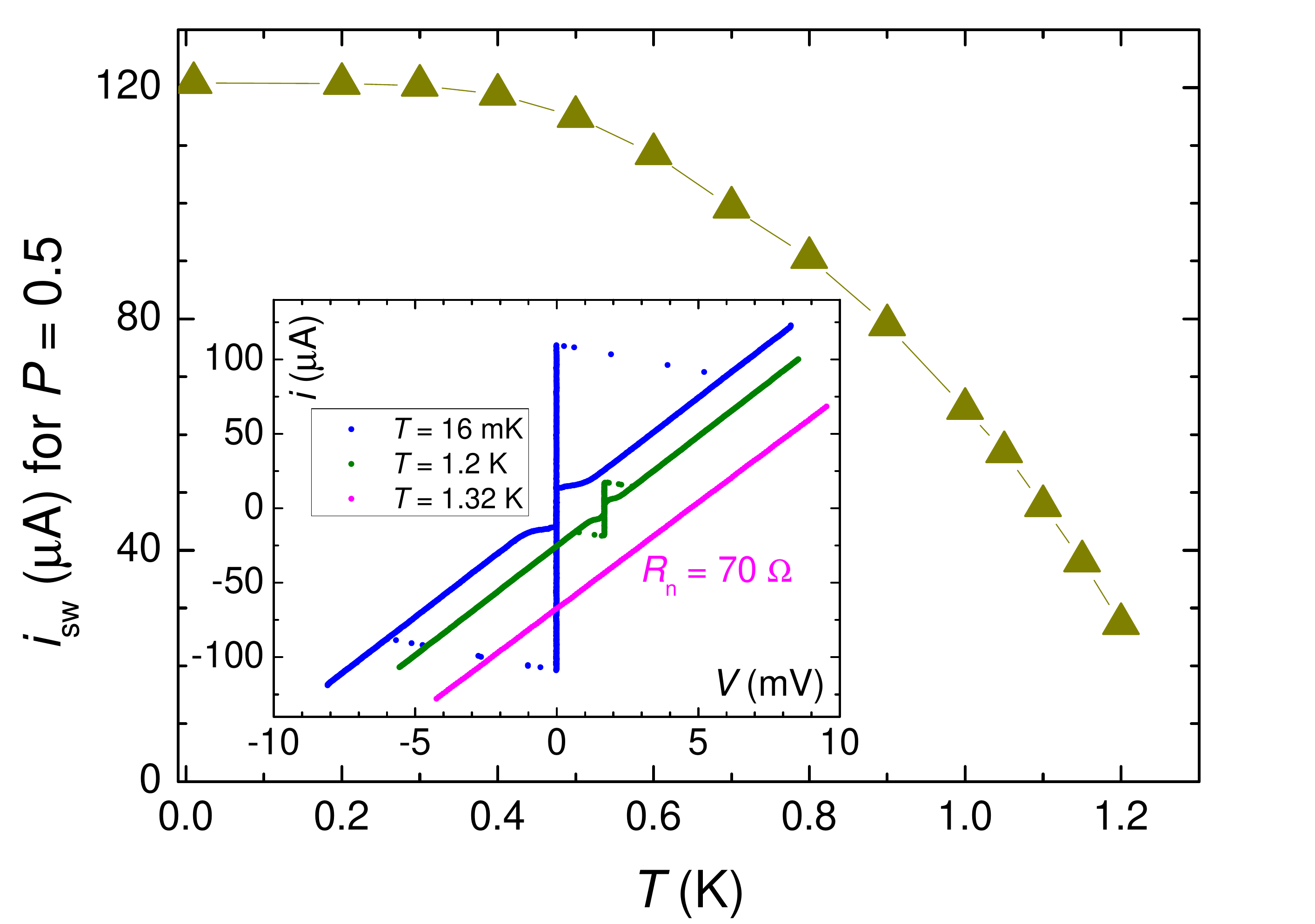}
\caption{\label{fig:Calibration}\textbf{Calibration curve.} Temperature dependence of the switching current for the superconducting weak link studied in this work. Inset: IV curves collected at three temperatures: 16.5\,mK, 1.2\,K, 1.32\,K (from left to right, offset horizontally).}
\end{figure}

\begin{figure*}
\centering
\includegraphics[width=1.0\textwidth]{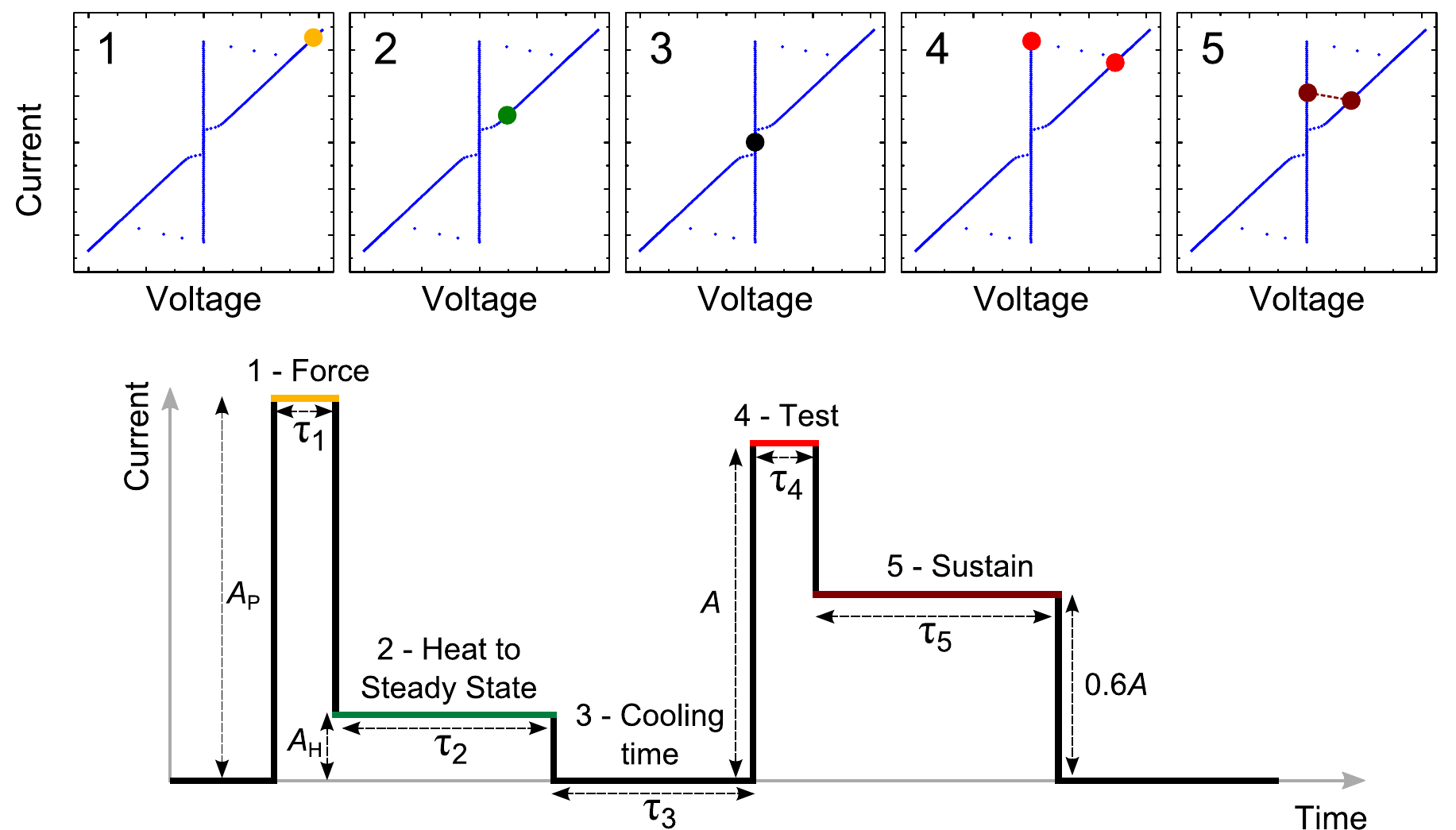}
\caption{\label{fig:Pumpandprobe}\textbf{The principle of the pump and probe experiment (\textit{pump}\&\textit{probe} \textit{pulse} definition).} By applying the current pulse larger than the switching threshold we force the junction to go to the normal state (1). Then we bring the junction and its surroundings to thermal steady state (2). The probe sequence (4,5) is delayed by time $\tau_3$ (3) with respect to the pump sequence (1,2) and its testing part (4), if tuned to obtain $P=0.5$ switching probability, measures the instantaneous temperature in the middle of the wire. The sustain part of the probe sequence (5) is to allow for read-out with slow room temperature electronics: if the junction switched a finite voltage is detected, otherwise no voltage builds up on the probing wires. $A_P$, $A_H$ and $A$ denote current amplitudes for different parts of the pump\&probe pulse.}
\end{figure*}

JJs are sometimes referred to as switches for their ability to carry supercurrent only to a certain level and, above this level, they switch to a finite voltage state (Fig.\,\ref{fig:Pulses}). The methodology of the switching current measurement is known\cite{Zgirski2011,DellaRocca2007,Zgirski2015,Chiorescu2003}. A rectangular current pulse is applied to the junction and the response of the junction is measured: it switches or remains in the superconducting state. The switching process exhibits stochastic character, for it involves thermal or quantum fluctuations driving JJ out of its metastable state\cite{Martinis1987,Devoret2003}. Sending a pulse train allows to determine the switching probability $P$ corresponding to a given pulse amplitude. Repeating the same experiment for different current amplitudes gives what is called S-curve: current amplitude dependence of the switching probability (Fig.\,\ref{fig:Pulses}). The switching experiments on JJs have shed some light on the nature of Andreev bound states in the superconducting point contacts\cite{Zgirski2011,DellaRocca2007}, allowed for magnetization measurements with nanoSQUIDs\cite{Wernsdorfer2009}, and have been statistically studied proving to be useful for generating random numbers\cite{Zgirski2015}. The key observation in the current context is the dependency of the switching current threshold on temperature (Fig.\,\ref{fig:Calibration})\cite{Dubos2001}, a feature required for a temperature sensor. The JJ thermometer is calibrated by measuring its switching current corresponding to $P = 0.5$ against the bath temperature $T_0$. To bring in the temporal resolution of the thermometer we make use of a \emph{pump}\&\emph{probe} idea, somewhat familiar from the laser physics. This is the key ingredient for our new approach: a nanostructure in thermal contact with the JJ is heated with a pump pulse and then, say several dozen of nanoseconds later, the JJ is tested with a probe pulse (Fig.\,\ref{fig:Pumpandprobe}). The probe pulse amplitude is adjusted to yield $P = 0.5$ switching probability. The delay between pump pulse and probe pulse can be controlled with accuracy of a single nanosecond, providing unprecedented resolution. It is worth to highlight a probe and hold feature of the JJ: the JJ reaches THz response bandwidth, but, due to hysteresis (retrapping current at which JJ returns to the superconducting state is much lower than the switching current) and with a properly tailored probe pulse (Fig.\,\ref{fig:Pumpandprobe}), read-out may be implemented with low frequency lines.

\subsection*{Model system for testing the proposed thermometry}
\begin{figure}
\centering
\includegraphics[width=0.5\textwidth]{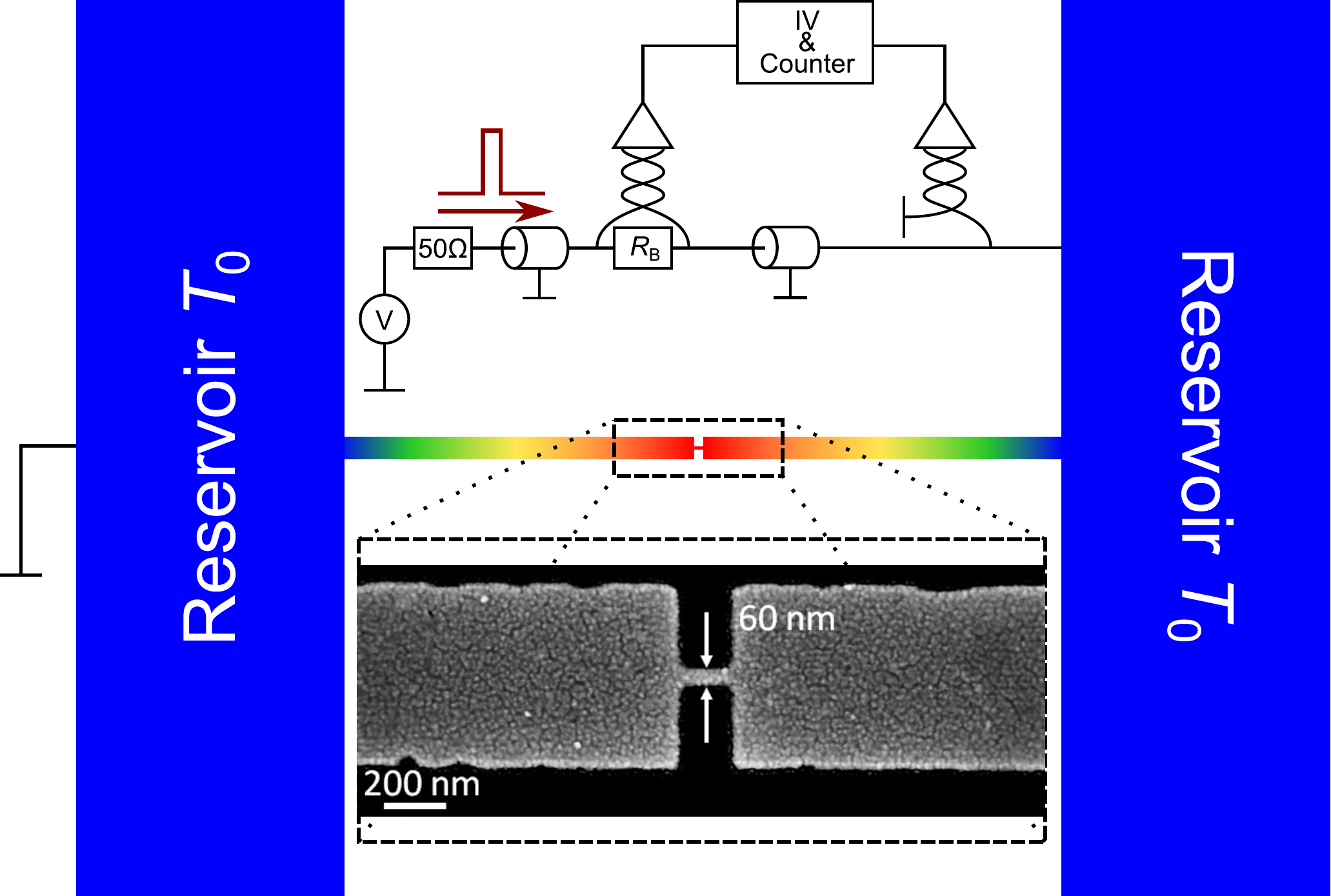}
\caption{\label{fig:Bridge}\textbf{Nanostructure and measurement setup used to benchmark the new thermometry.} The wire of length 75\,$\mu$m is interrupted in the middle with a Dayem nanobridge. The width of the wire is 600\,nm and its thickness is 30\,nm. Two voltage amplifiers depicted with triangles measure current $i_J$ flowing into nanobridge and voltage $V_J$ across it.}
\end{figure}

\begin{figure}
\centering
\includegraphics[width=0.5\textwidth]{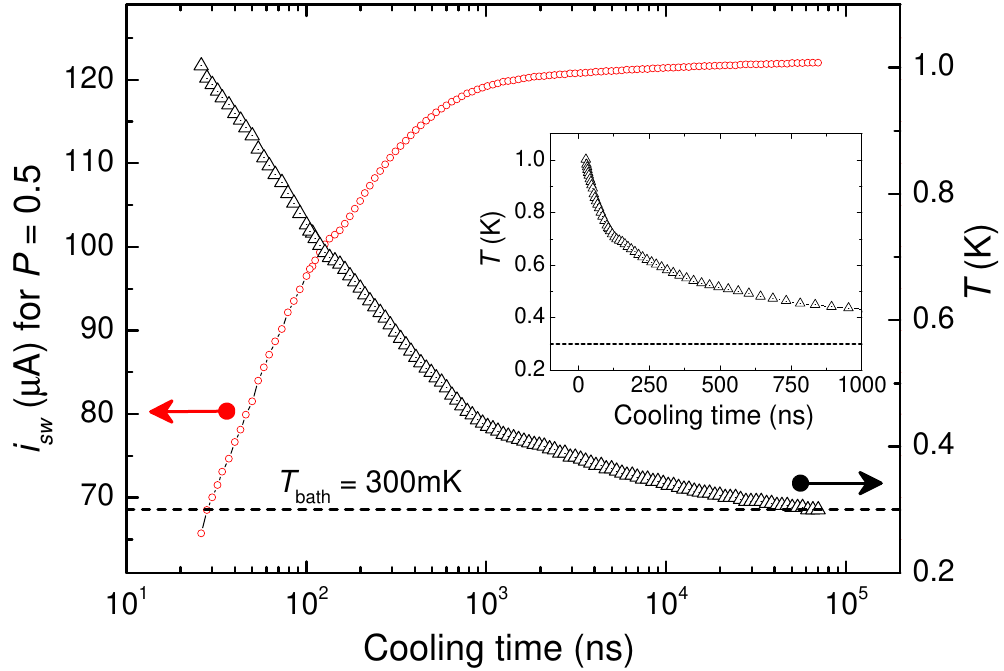}
\caption{\label{fig:Relaxation}\textbf{Nanowire thermal relaxation.} Switching current temporal changes recorded for the bridge undergoing thermal relaxation to the bath temperature of 300\,mK (circles) and the corresponding electron temperature inferred with the aid of the $i_{sw}(T)$ curve (triangles). A graphical conversion is presented in the Supplementary Note 4. The pump\&probe pulse used in the experiment is shown in the Fig.\,\ref{fig:Pumpandprobe}. The timing is: $\tau_1$ = 100\,ns, $\tau_2$ = 5\,$\mu$s, $\tau_3$ = cooling time, $\tau_4$ = 10\,ns, $\tau_5$ = 3\,$\mu$s. Inset shows the same $T$ vs. cooling time data in the linear scale.}
\end{figure}

The suggested thermometry scheme can be realized based on different types of JJs, such as tunneling JJs (a very thin oxide layer sandwiched between two superconducting electrodes), proximity JJs (a piece of normal metal interrupting superconductor) or superconducting bridges e.g. Dayem nanobridges, etc.\cite{Tinkham2004}. In the current work the Aluminum Dayem nanobridge has been utilized to demonstrate applicability and reliability of the new thermometry and highlight its superior temporal sensitivity. The device is presented in Fig.\,\ref{fig:Bridge}. It consists of a narrow superconducting bridge placed in the middle of superconducting wire anchored at both ends to large area contact pads serving as energy reservoirs. Such a structure is easily obtained on a silicon substrate with conventional one-step e-beam lithography and, what is important for benchmarking, its thermal dynamics is easy to simulate as thermal properties of Aluminum are well known. The device is placed in the dilution refrigerator with base temperature of 10\,mK. First we measure its switching current dependence on temperature $i_{sw}(T)$ at well defined bath temperatures determined with a conventional calibrated RuO$_x$ thermometer (Fig.\,\ref{fig:Calibration}). Then, with application of pump\&probe pulse train, we perform switching current relaxation measurements of the junction after it switched first to a normal state, was then brought to a steady state and was finally left to cool down. For each delay between pump pulse and probe pulse, we find the switching current amplitude corresponding to $P = 0.5$ switching probability. We have sent train of $N=10,000$ pump\&probe pulses to measure each point. Period of 100\,$\mu$sec have ensured complete thermal relaxation after each pump\&probe pulse. Switching current temporal changes for the junction heated above a critical current and relaxing back to the bath temperature of 300\,mK are displayed in Fig.\,\ref{fig:Relaxation}. To convert switching current into dynamic temperature we use the calibration relationship $i_{sw}(T)$ (Fig.\,\ref{fig:Calibration}). Such an approach imposes the maximum temporal uncertainty of the temperature determination equal to the testing pulse duration i.e. about 10\,ns in the presented study (see the Supplementary Note 3), but allows for a straightforward use of the calibration curve $i_{sw}(T)$ for recalculating of the switching current into dynamic temperature. We highlight an outstanding temporal resolution using logarithmic scale. It reveals nanosecond resolving capability of the thermometer. We can monitor temperature of the link immediately after it reenters to superconducting state, about 20\,ns after switching off the heating current (Fig.\,\ref{fig:Relaxation}). This time is refered to as the dead time of our experiment. We use the same weak link both for heating the wire to an elevated temperature and for sensing the dynamic temperature during relaxation. For heating we need to transfer the weak link to the normal state, for sensing it must be in the superconducting state. Transition between these two regimes requires several dozen of nanoseconds. It is possible to make these two functionalities independent by introducing a separate heater and eliminate dead time in experiments.

\subsection*{Steady state temperature profile - Modeling}
\begin{figure}
\centering
\includegraphics[width=0.5\textwidth]{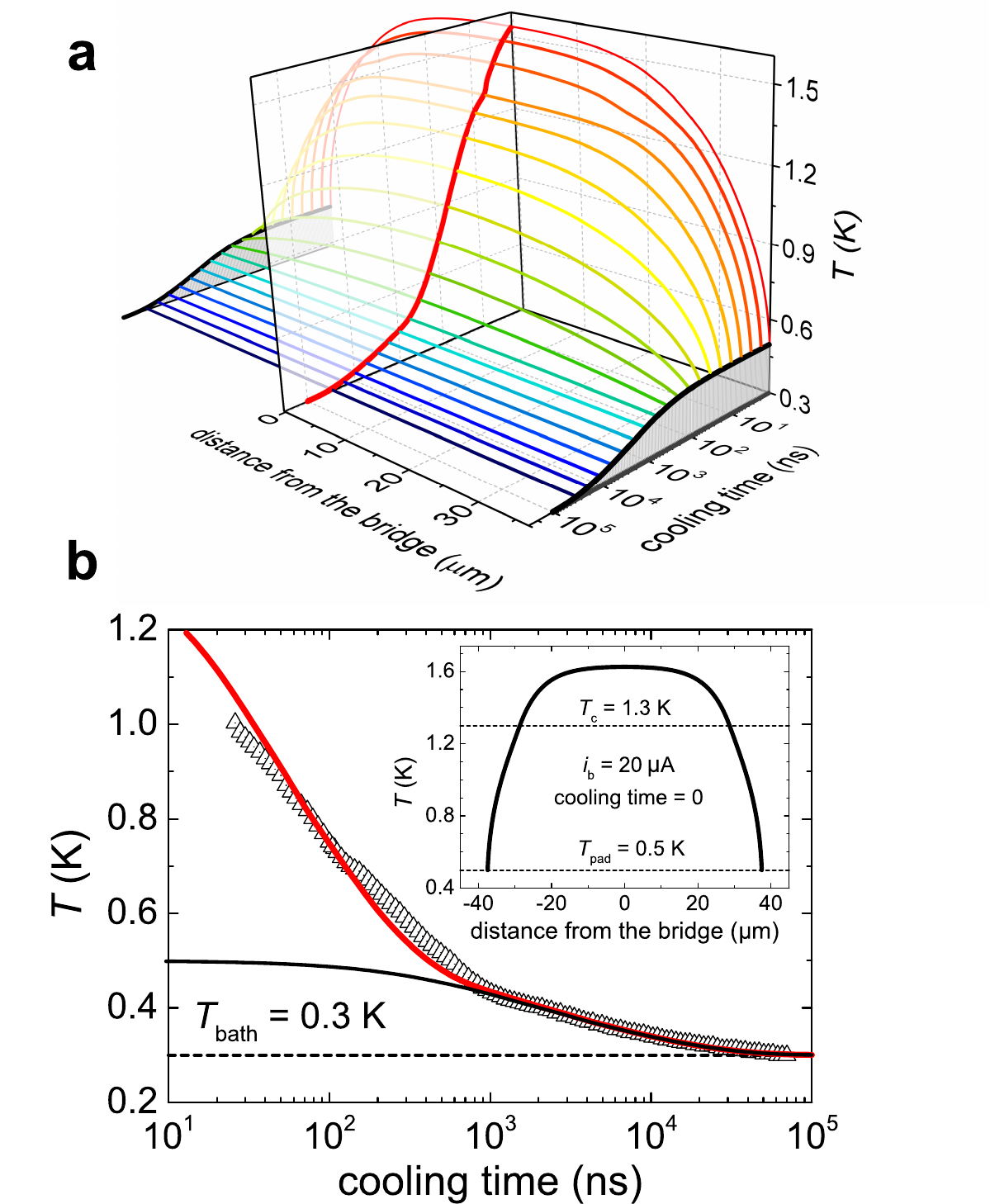}
\caption{\label{fig:Solution}\textbf{Numerical modeling of temperature relaxation in the superconducting nanowire.} \textbf{(a)} Temporal evolution of temperature profile in the wire. Temperature variations in the bridge and in the pads are distinguished with separate curves (red and black respectively). \textbf{(b)} Modeled bridge (red solid curve) and pad (black solid curve) temperatures (the same as distinguished curves from panel (a)) compared with experimental relaxation (black triangles, the same data as in Fig.\,\ref{fig:Relaxation}). Cooling time $t=0$ is set to be the end of the heating pulse bringing the wire to a steady state with $T\simeq 1.6$\,K in the center of the wire (at the nanobridge location). The corresponding temperature profile is shown in the inset.}
\end{figure}

To get a temperature profile of the wire once the wire switches to the normal state we solve the heat balance equation in a steady state characterized by a constant electric current dissipating the Joule heat in the wire. The equation reads:
\begin{equation}
-\cfrac{d}{dx}\left(\kappa(T_e)\cfrac{dT_e}{dx}\right)=f(T_e)
\end{equation}
where left part of equation deals with hot electron diffusion ($\kappa(T_e)$ is the electron thermal conductivity) and $f(T_e)=H(T_c)\cdot\cfrac{r\cdot i_b^2}{S}-\dot{q}_{ep}(T_e)$ is a source/drain term accounting for the heat generation and absorption in a unit length of the wire with $r$ - resistance per unit length, S - wire cross-section, $i_b$ - biasing current, $H(T_c)$ - the Heaviside step function and $\dot{q}_{ep}(T_e)$ - hot electron power fed back to phonons. $\kappa(T_e)$ and $\dot{q}_{ep}(T_e)$ are numerical data calculated according to the integrals found in ref.\,\onlinecite{Courtois2008} and \,\onlinecite{Maisi2013}. At $T_e>T_c$, $\dot{q}_{ep}(T_e)$ is equal to $\sum (T_e^5-T_{ph}^5)$ with $\sum = 1.8 \times 10^9\,$W/m$^3/$K$^5$ being the electron-phonon coupling constant in Aluminum. The theoretical steady state profile corresponding to our experimental realization is presented in the inset of Fig.\,\ref{fig:Solution}. Since the profile is flat in the center of the wire the only heat transfer under consideration is heat flow from hot electrons to phonons $(f(T_e) = 0)$, yielding for our experiment with $A_H = 20$\,$\mu$A (cf. Fig.\,\ref{fig:Pumpandprobe}) electron temperature of $T_e \simeq 1.6$\,K in the center of the wire (at the nanobridge location).

\subsection*{Relaxation - Modeling}
The cooling of the wire is governed by the temporal relaxation equation taking into account two relaxation paths for the excess electron energy: the electron-phonon coupling and the diffusion of hot electrons:
\begin{equation}
-\cfrac{d}{dx}\left(\kappa(T_e)\cfrac{dT_e}{dx}\right)=c_s(T_e)\cdot \cfrac{\delta T_e}{\delta t}+\dot{q}_{ep}(T_e)
\end{equation}
where $c_s(T_e)$ is experimentally determined heat capacity of Aluminum found in ref.\,\onlinecite{Phillips1959}. The initial condition is assumed to be the steady state profile introduced above and displayed in the inset of Fig.\,\ref{fig:Solution}. The calculated temperature relaxation in the superconducting wire and the measurement results are presented in Fig.\,\ref{fig:Solution}. The fast relaxation within first microsecond involves processes leading to equalizing the temperature in the whole wire with the temperature of the pads. More detailed analysis (see the Supplementary Note 6) shows that above $\simeq$ 0.9\,K electron-phonon relaxation channel dominates while below $\simeq$ 0.6\,K hot electron diffusion takes over as electron-phonon becomes very inefficient. The slow relaxation reflects dynamics of the electron-phonon heat transfer in the pads. On the assumption that the pads are perfect energy reservoirs the model remains in qualitative agreement with experimental relaxation profile during first microsecond but it is not able to explain the long relaxation tail extending up into tens of microsecond range. Similar modeling for initial temperature of the center of the wire equal to 3.3\,K is presented in the Supplementary Note 7. It is important to notice that a small adjustment in the heat capacity and electron-phonon coupling data may account for the observed difference between experimental points and theory. Since our primary goal is to demonstrate the new approach to thermometry and its temporal resolution superiority, rather than to determine detailed temperature dependence of the Aluminum heat capacity, we have calculated the temperature relaxation without any adjustment to material parameters basing solely on the data available from literature.

\subsection*{Discussion}
We have benchmarked the ultrafast thermometry protocol employing an Aluminum Dayem nanobridge as a temperature-sensing element. However the method will work for any kind of JJ which exhibits switching current dependence on temperature. Subject to the temperature interval of interest one may use a tunnel junction or proximity junction e.g. superconductor-normal-metal-superconductor SNS junction. By selection of proper materials and adjusting the length of the normal bridge (in the latter case) one can engineer operational temperature range and magnitude of the switching current to match experimental requirements.

Temperature is an equilibrium concept and its application to describe a state of rapidly heating or cooling system may rise some questions. In solid state physics temperature determines the occupation of electron states (in equilibrium according to the Fermi-Dirac distribution). There are situations when electron distribution is far from equilibrium, but without external excitation it locally quickly converges to the Fermi-Dirac function due to fast electron-electron interaction. It happens at time scales much shorter than relaxation times considered in the presented experiment. We assume that electron subsystem undergoes quasi-static evolution and is described locally at each moment with equilibrium Fermi-Dirac distribution. Thus temperature of electron system is well-defined all the time during relaxation. Our concept of temperature measurement can be also extended to probe non-equilibrium systems i.e. where electrons are not Fermi-Dirac distributed. In such a case the measured temperature is called \textit{effective temperature} and corresponds to the equilibrium temperature if the same physical effect is observed e.g. the same switching current.

Variations of switching current in JJs are set by two factors: (i) value of the superconducting gap (it is local property of the junction that sets the critical current) and (ii) strength of electromagnetic fluctuations induced by environment (it is non-local property that may make junction sensitive to temperature of remote impedance). In case of a Dayem nanobridge with critical current of about 100\,$\mu$A, as the one studied here, first factor dominates and fluctuations play a minor role. However, for a tunnel junction with low critical current fluctuations will dominate and a sensor based on such a junction will be sensitive to a non-local temperature of the environment.

The presented thermometry is unique with respect to power dissipation. Pulses probing JJ thermometer have steep rising slopes and with standard equipment can be made shorter than 1\,ns. Consequently, they cause negligible heating of the sample and do not rise sample temperature. It is in contrast to familiar DC and RF techniques where probing signals may alter temperature which one wishes to measure.

\subsection*{Outlook}
The proposed thermometer can be considered as a base sensing element for bolometers operating in the far-infrared, THz and microwave bands. Owing to its very small size the JJ-based thermometer could be integrated with very small absorber with all three dimensions falling into a few tens of nanometers providing a versatile platform for sub-aJ/K calorimetry. Small metal volume of 10$^5$-10$^6$\,nm$^3$ employed as the absorber yields a heat capacity of $(10^2 - 10^3)k_B$ offering high gains in sensitivity and temporal resolution of incident radiation\cite{note}. A single microwave photon absorption (2-20\,GHz respectively) would produce a detectable 10\,mK temperature spike, allowing to count arriving photons and investigate statistics of heat transport in superconducting quantum circuits. The proposed thermometers, defined in the various places of the electronic chip, will offer fast and high spatial resolution for mapping the temperature across the chip. They will allow to analyze the effect of fast electrical pulses on the chip temperature distribution thus helping to develop cryoelectronics and quantum computing devices. The unprecedented temporal resolution of the thermometer allows to "see" heat propagating across different nanostructures e.g. get access to real time observation of hot electron diffusion and investigate mechanisms of heat dissipation via phonon or photon emission channels\cite{Pekola2006}. The temporal resolution of thermometer can be pushed down into sub-ns range with application of standard GHz-limited arbitrary waveform generators and sample design compatible with microwave propagation.
\vspace{1cm}
\subsection*{Conclusion}
We have presented implementation of the hysteretic superconducting weak link for nanosecond thermometry applications. Such a weak link exhibits a very fast intrinsic dynamics falling into picoseconds range and is perfectly suited for sensing rapidly changing electron temperature, with the temporal resolution of a single nanosecond easily achievable. Successful implementation of our approach paves the way to cutting-edge experiments in the field of thermodynamics of low temperature quantum circuits. It gives rise to development of ultra-sensitive low noise calorimeters and bolometers for health, security and astronomical applications.

\section*{Methods}
The measurements were performed in the Triton 400 dilution refrigerator. Pulses for testing the weak link were tailored with Agilent 33250A AWG and sent down the junction with coaxial line equipped with attenuators at different temperature stages. Switching events were recorded and counted with LeCroy HRO 66Zi oscilloscope and Philips PM 6680 counter.
The steady state temperature profiles in the wire were calculated with the Runge Kutta method of 4$^{th}$ order treating the two coupled 1$^{st}$ order differential equations. The heat flow equation was solved with Matlab Partial Differential Equation Toolbox.

\bibliography{Nanosecond_thermometry}

\begin{thebibliography}{29}%
\makeatletter
\providecommand \@ifxundefined [1]{%
 \@ifx{#1\undefined}
}%
\providecommand \@ifnum [1]{%
 \ifnum #1\expandafter \@firstoftwo
 \else \expandafter \@secondoftwo
 \fi
}%
\providecommand \@ifx [1]{%
 \ifx #1\expandafter \@firstoftwo
 \else \expandafter \@secondoftwo
 \fi
}%
\providecommand \natexlab [1]{#1}%
\providecommand \enquote  [1]{``#1''}%
\providecommand \bibnamefont  [1]{#1}%
\providecommand \bibfnamefont [1]{#1}%
\providecommand \citenamefont [1]{#1}%
\providecommand \href@noop [0]{\@secondoftwo}%
\providecommand \href [0]{\begingroup \@sanitize@url \@href}%
\providecommand \@href[1]{\@@startlink{#1}\@@href}%
\providecommand \@@href[1]{\endgroup#1\@@endlink}%
\providecommand \@sanitize@url [0]{\catcode `\\12\catcode `\$12\catcode
  `\&12\catcode `\#12\catcode `\^12\catcode `\_12\catcode `\%12\relax}%
\providecommand \@@startlink[1]{}%
\providecommand \@@endlink[0]{}%
\providecommand \url  [0]{\begingroup\@sanitize@url \@url }%
\providecommand \@url [1]{\endgroup\@href {#1}{\urlprefix }}%
\providecommand \urlprefix  [0]{URL }%
\providecommand \Eprint [0]{\href }%
\providecommand \doibase [0]{http://dx.doi.org/}%
\providecommand \selectlanguage [0]{\@gobble}%
\providecommand \bibinfo  [0]{\@secondoftwo}%
\providecommand \bibfield  [0]{\@secondoftwo}%
\providecommand \translation [1]{[#1]}%
\providecommand \BibitemOpen [0]{}%
\providecommand \bibitemStop [0]{}%
\providecommand \bibitemNoStop [0]{.\EOS\space}%
\providecommand \EOS [0]{\spacefactor3000\relax}%
\providecommand \BibitemShut  [1]{\csname bibitem#1\endcsname}%
\let\auto@bib@innerbib\@empty
\bibitem [{\citenamefont {Gasparinetti}\ \emph {et~al.}(2015)\citenamefont
  {Gasparinetti}, \citenamefont {Viisanen}, \citenamefont {Saira},
  \citenamefont {Faivre}, \citenamefont {Arzeo}, \citenamefont {Meschke},\ and\
  \citenamefont {Pekola}}]{Gasparinetti2015}%
  \BibitemOpen
  \bibfield  {author} {\bibinfo {author} {\bibfnamefont {S.}~\bibnamefont
  {Gasparinetti}}, \bibinfo {author} {\bibfnamefont {K.~L.}\ \bibnamefont
  {Viisanen}}, \bibinfo {author} {\bibfnamefont {O.-P.}\ \bibnamefont {Saira}},
  \bibinfo {author} {\bibfnamefont {T.}~\bibnamefont {Faivre}}, \bibinfo
  {author} {\bibfnamefont {M.}~\bibnamefont {Arzeo}}, \bibinfo {author}
  {\bibfnamefont {M.}~\bibnamefont {Meschke}}, \ and\ \bibinfo {author}
  {\bibfnamefont {J.~P.}\ \bibnamefont {Pekola}},\ }\bibfield  {title}
  {\enquote {\bibinfo {title} {Fast electron thermometry for ultrasensitive
  calorimetric detection},}\ }\href {\doibase 10.1103/PhysRevApplied.3.014007}
  {\bibfield  {journal} {\bibinfo  {journal} {Phys. Rev. Applied}\ }\textbf
  {\bibinfo {volume} {3}},\ \bibinfo {pages} {014007} (\bibinfo {year}
  {2015})}\BibitemShut {NoStop}%
\bibitem [{\citenamefont {Schmidt}\ \emph {et~al.}(2004)\citenamefont
  {Schmidt}, \citenamefont {Yung},\ and\ \citenamefont
  {Cleland}}]{Cleland2004}%
  \BibitemOpen
  \bibfield  {author} {\bibinfo {author} {\bibfnamefont {D.~R.}\ \bibnamefont
  {Schmidt}}, \bibinfo {author} {\bibfnamefont {C.~S.}\ \bibnamefont {Yung}}, \
  and\ \bibinfo {author} {\bibfnamefont {A.~N.}\ \bibnamefont {Cleland}},\
  }\bibfield  {title} {\enquote {\bibinfo {title} {Temporal measurement of
  hot-electron relaxation in a phonon-cooled metal island},}\ }\href {\doibase
  10.1103/PhysRevB.69.140301} {\bibfield  {journal} {\bibinfo  {journal} {Phys.
  Rev. B}\ }\textbf {\bibinfo {volume} {69}},\ \bibinfo {pages} {140301}
  (\bibinfo {year} {2004})}\BibitemShut {NoStop}%
\bibitem [{\citenamefont {Saira}\ \emph {et~al.}(2016)\citenamefont {Saira},
  \citenamefont {Zgirski}, \citenamefont {Viisanen}, \citenamefont {Golubev},\
  and\ \citenamefont {Pekola}}]{Pekola2016}%
  \BibitemOpen
  \bibfield  {author} {\bibinfo {author} {\bibfnamefont {O.-P.}\ \bibnamefont
  {Saira}}, \bibinfo {author} {\bibfnamefont {M.}~\bibnamefont {Zgirski}},
  \bibinfo {author} {\bibfnamefont {K.~L.}\ \bibnamefont {Viisanen}}, \bibinfo
  {author} {\bibfnamefont {D.~S.}\ \bibnamefont {Golubev}}, \ and\ \bibinfo
  {author} {\bibfnamefont {J.~P.}\ \bibnamefont {Pekola}},\ }\bibfield  {title}
  {\enquote {\bibinfo {title} {Dispersive thermometry with a {J}osephson
  junction coupled to a resonator},}\ }\href {\doibase
  10.1103/PhysRevApplied.6.024005} {\bibfield  {journal} {\bibinfo  {journal}
  {Phys. Rev. Applied}\ }\textbf {\bibinfo {volume} {6}},\ \bibinfo {pages}
  {024005} (\bibinfo {year} {2016})}\BibitemShut {NoStop}%
\bibitem [{\citenamefont {Feshchenko}\ \emph {et~al.}(2015)\citenamefont
  {Feshchenko}, \citenamefont {Casparis}, \citenamefont {Khaymovich},
  \citenamefont {Maradan}, \citenamefont {Saira}, \citenamefont {Palma},
  \citenamefont {Meschke}, \citenamefont {Pekola},\ and\ \citenamefont
  {Zumb\"uhl}}]{Feshchenko2015}%
  \BibitemOpen
  \bibfield  {author} {\bibinfo {author} {\bibfnamefont {A.~V.}\ \bibnamefont
  {Feshchenko}}, \bibinfo {author} {\bibfnamefont {L.}~\bibnamefont
  {Casparis}}, \bibinfo {author} {\bibfnamefont {I.~M.}\ \bibnamefont
  {Khaymovich}}, \bibinfo {author} {\bibfnamefont {D.}~\bibnamefont {Maradan}},
  \bibinfo {author} {\bibfnamefont {O.-P.}\ \bibnamefont {Saira}}, \bibinfo
  {author} {\bibfnamefont {M.}~\bibnamefont {Palma}}, \bibinfo {author}
  {\bibfnamefont {M.}~\bibnamefont {Meschke}}, \bibinfo {author} {\bibfnamefont
  {J.~P.}\ \bibnamefont {Pekola}}, \ and\ \bibinfo {author} {\bibfnamefont
  {D.~M.}\ \bibnamefont {Zumb\"uhl}},\ }\bibfield  {title} {\enquote {\bibinfo
  {title} {Tunnel-junction thermometry down to millikelvin temperatures},}\
  }\href {\doibase 10.1103/PhysRevApplied.4.034001} {\bibfield  {journal}
  {\bibinfo  {journal} {Phys. Rev. Applied}\ }\textbf {\bibinfo {volume} {4}},\
  \bibinfo {pages} {034001} (\bibinfo {year} {2015})}\BibitemShut {NoStop}%
\bibitem [{\citenamefont {Beyer}\ \emph {et~al.}(2007)\citenamefont {Beyer},
  \citenamefont {Drung}, \citenamefont {Kirste}, \citenamefont {Engert},
  \citenamefont {Netsch}, \citenamefont {Fleischmann},\ and\ \citenamefont
  {Enss}}]{Beyer2007}%
  \BibitemOpen
  \bibfield  {author} {\bibinfo {author} {\bibfnamefont {J.}~\bibnamefont
  {Beyer}}, \bibinfo {author} {\bibfnamefont {D.}~\bibnamefont {Drung}},
  \bibinfo {author} {\bibfnamefont {A.}~\bibnamefont {Kirste}}, \bibinfo
  {author} {\bibfnamefont {J.}~\bibnamefont {Engert}}, \bibinfo {author}
  {\bibfnamefont {A.}~\bibnamefont {Netsch}}, \bibinfo {author} {\bibfnamefont
  {A.}~\bibnamefont {Fleischmann}}, \ and\ \bibinfo {author} {\bibfnamefont
  {C.}~\bibnamefont {Enss}},\ }\bibfield  {title} {\enquote {\bibinfo {title}
  {A magnetic-field-fluctuation thermometer for the m{K} range based on
  {SQUID}-magnetometry},}\ }\href {\doibase 10.1109/TASC.2007.898265}
  {\bibfield  {journal} {\bibinfo  {journal} {IEEE Transactions on Applied
  Superconductivity}\ }\textbf {\bibinfo {volume} {17}},\ \bibinfo {pages}
  {760--763} (\bibinfo {year} {2007})}\BibitemShut {NoStop}%
\bibitem [{\citenamefont {Gasparinetti}\ \emph {et~al.}(2011)\citenamefont
  {Gasparinetti}, \citenamefont {Deon}, \citenamefont {Biasiol}, \citenamefont
  {Sorba}, \citenamefont {Beltram},\ and\ \citenamefont
  {Giazotto}}]{Gasparinetti2011}%
  \BibitemOpen
  \bibfield  {author} {\bibinfo {author} {\bibfnamefont {S.}~\bibnamefont
  {Gasparinetti}}, \bibinfo {author} {\bibfnamefont {F.}~\bibnamefont {Deon}},
  \bibinfo {author} {\bibfnamefont {G.}~\bibnamefont {Biasiol}}, \bibinfo
  {author} {\bibfnamefont {L.}~\bibnamefont {Sorba}}, \bibinfo {author}
  {\bibfnamefont {F.}~\bibnamefont {Beltram}}, \ and\ \bibinfo {author}
  {\bibfnamefont {F.}~\bibnamefont {Giazotto}},\ }\bibfield  {title} {\enquote
  {\bibinfo {title} {Probing the local temperature of a two-dimensional
  electron gas microdomain with a quantum dot: Measurement of electron-phonon
  interaction},}\ }\href {\doibase 10.1103/PhysRevB.83.201306} {\bibfield
  {journal} {\bibinfo  {journal} {Phys. Rev. B}\ }\textbf {\bibinfo {volume}
  {83}},\ \bibinfo {pages} {201306} (\bibinfo {year} {2011})}\BibitemShut
  {NoStop}%
\bibitem [{\citenamefont {Wellstood}\ \emph {et~al.}(1994)\citenamefont
  {Wellstood}, \citenamefont {Urbina},\ and\ \citenamefont
  {Clarke}}]{Clarke1994}%
  \BibitemOpen
  \bibfield  {author} {\bibinfo {author} {\bibfnamefont {F.~C.}\ \bibnamefont
  {Wellstood}}, \bibinfo {author} {\bibfnamefont {C.}~\bibnamefont {Urbina}}, \
  and\ \bibinfo {author} {\bibfnamefont {John}\ \bibnamefont {Clarke}},\
  }\bibfield  {title} {\enquote {\bibinfo {title} {Hot-electron effects in
  metals},}\ }\href {\doibase 10.1103/PhysRevB.49.5942} {\bibfield  {journal}
  {\bibinfo  {journal} {Phys. Rev. B}\ }\textbf {\bibinfo {volume} {49}},\
  \bibinfo {pages} {5942--5955} (\bibinfo {year} {1994})}\BibitemShut {NoStop}%
\bibitem [{\citenamefont {Schwab}\ \emph {et~al.}(2000)\citenamefont {Schwab},
  \citenamefont {Henriksen}, \citenamefont {Worlock},\ and\ \citenamefont
  {Roukes}}]{Schwab2000}%
  \BibitemOpen
  \bibfield  {author} {\bibinfo {author} {\bibfnamefont {K.}~\bibnamefont
  {Schwab}}, \bibinfo {author} {\bibfnamefont {E.~A.}\ \bibnamefont
  {Henriksen}}, \bibinfo {author} {\bibfnamefont {J.~M.}\ \bibnamefont
  {Worlock}}, \ and\ \bibinfo {author} {\bibfnamefont {M.~L.}\ \bibnamefont
  {Roukes}},\ }\bibfield  {title} {\enquote {\bibinfo {title} {Measurement of
  the quantum of thermal conductance},}\ }\href {\doibase 10.1038/35010065}
  {\bibfield  {journal} {\bibinfo  {journal} {Nature}\ }\textbf {\bibinfo
  {volume} {404}},\ \bibinfo {pages} {974--977} (\bibinfo {year}
  {2000})}\BibitemShut {NoStop}%
\bibitem [{\citenamefont {Meschke}\ \emph {et~al.}(2006)\citenamefont
  {Meschke}, \citenamefont {Guichard},\ and\ \citenamefont
  {Pekola}}]{Pekola2006}%
  \BibitemOpen
  \bibfield  {author} {\bibinfo {author} {\bibfnamefont {M.}~\bibnamefont
  {Meschke}}, \bibinfo {author} {\bibfnamefont {W.}~\bibnamefont {Guichard}}, \
  and\ \bibinfo {author} {\bibfnamefont {J.~P.}\ \bibnamefont {Pekola}},\
  }\bibfield  {title} {\enquote {\bibinfo {title} {Single-mode heat conduction
  by photons},}\ }\href {\doibase 10.1038/nature05276} {\bibfield  {journal}
  {\bibinfo  {journal} {Nature}\ }\textbf {\bibinfo {volume} {444}},\ \bibinfo
  {pages} {187--190} (\bibinfo {year} {2006})}\BibitemShut {NoStop}%
\bibitem [{\citenamefont {Jezouin}\ \emph {et~al.}(2013)\citenamefont
  {Jezouin}, \citenamefont {Parmentier}, \citenamefont {Anthore}, \citenamefont
  {Gennser}, \citenamefont {Cavanna}, \citenamefont {Jin},\ and\ \citenamefont
  {Pierre}}]{Jezouin2013}%
  \BibitemOpen
  \bibfield  {author} {\bibinfo {author} {\bibfnamefont {S.}~\bibnamefont
  {Jezouin}}, \bibinfo {author} {\bibfnamefont {F.~D.}\ \bibnamefont
  {Parmentier}}, \bibinfo {author} {\bibfnamefont {A.}~\bibnamefont {Anthore}},
  \bibinfo {author} {\bibfnamefont {U.}~\bibnamefont {Gennser}}, \bibinfo
  {author} {\bibfnamefont {A.}~\bibnamefont {Cavanna}}, \bibinfo {author}
  {\bibfnamefont {Y.}~\bibnamefont {Jin}}, \ and\ \bibinfo {author}
  {\bibfnamefont {F.}~\bibnamefont {Pierre}},\ }\bibfield  {title} {\enquote
  {\bibinfo {title} {Quantum limit of heat flow across a single electronic
  channel},}\ }\href {\doibase 10.1126/science.1241912} {\bibfield  {journal}
  {\bibinfo  {journal} {Science}\ }\textbf {\bibinfo {volume} {342}},\ \bibinfo
  {pages} {601--604} (\bibinfo {year} {2013})}\BibitemShut {NoStop}%
\bibitem [{\citenamefont {Koski}\ \emph {et~al.}(2015)\citenamefont {Koski},
  \citenamefont {Kutvonen}, \citenamefont {Khaymovich}, \citenamefont
  {Ala-Nissila},\ and\ \citenamefont {Pekola}}]{Koski2015}%
  \BibitemOpen
  \bibfield  {author} {\bibinfo {author} {\bibfnamefont {J.~V.}\ \bibnamefont
  {Koski}}, \bibinfo {author} {\bibfnamefont {A.}~\bibnamefont {Kutvonen}},
  \bibinfo {author} {\bibfnamefont {I.~M.}\ \bibnamefont {Khaymovich}},
  \bibinfo {author} {\bibfnamefont {T.}~\bibnamefont {Ala-Nissila}}, \ and\
  \bibinfo {author} {\bibfnamefont {J.~P.}\ \bibnamefont {Pekola}},\ }\bibfield
   {title} {\enquote {\bibinfo {title} {On-chip maxwell's demon as an
  information-powered refrigerator},}\ }\href {\doibase
  10.1103/PhysRevLett.115.260602} {\bibfield  {journal} {\bibinfo  {journal}
  {Phys. Rev. Lett.}\ }\textbf {\bibinfo {volume} {115}},\ \bibinfo {pages}
  {260602} (\bibinfo {year} {2015})}\BibitemShut {NoStop}%
\bibitem [{\citenamefont {Nahum}\ \emph {et~al.}(1994)\citenamefont {Nahum},
  \citenamefont {Eiles},\ and\ \citenamefont {Martinis}}]{Nahum1994}%
  \BibitemOpen
  \bibfield  {author} {\bibinfo {author} {\bibfnamefont {M.}~\bibnamefont
  {Nahum}}, \bibinfo {author} {\bibfnamefont {T.~M.}\ \bibnamefont {Eiles}}, \
  and\ \bibinfo {author} {\bibfnamefont {John~M.}\ \bibnamefont {Martinis}},\
  }\bibfield  {title} {\enquote {\bibinfo {title} {Electronic microrefrigerator
  based on a normal-insulator-superconductor tunnel junction},}\ }\href
  {\doibase 10.1063/1.112456} {\bibfield  {journal} {\bibinfo  {journal}
  {Applied Physics Letters}\ }\textbf {\bibinfo {volume} {65}},\ \bibinfo
  {pages} {3123--3125} (\bibinfo {year} {1994})}\BibitemShut {NoStop}%
\bibitem [{\citenamefont {Schmidt}\ \emph {et~al.}(2003)\citenamefont
  {Schmidt}, \citenamefont {Yung},\ and\ \citenamefont
  {Cleland}}]{Cleland2003}%
  \BibitemOpen
  \bibfield  {author} {\bibinfo {author} {\bibfnamefont {D.~R.}\ \bibnamefont
  {Schmidt}}, \bibinfo {author} {\bibfnamefont {C.~S.}\ \bibnamefont {Yung}}, \
  and\ \bibinfo {author} {\bibfnamefont {A.~N.}\ \bibnamefont {Cleland}},\
  }\bibfield  {title} {\enquote {\bibinfo {title} {Nanoscale radio-frequency
  thermometry},}\ }\href {\doibase 10.1063/1.1597983} {\bibfield  {journal}
  {\bibinfo  {journal} {Applied Physics Letters}\ }\textbf {\bibinfo {volume}
  {83}},\ \bibinfo {pages} {1002--1004} (\bibinfo {year} {2003})}\BibitemShut
  {NoStop}%
\bibitem [{\citenamefont {Fitzgerald}\ \emph {et~al.}(2006)\citenamefont
  {Fitzgerald}, \citenamefont {Wallace}, \citenamefont {Jimenez-Linan},
  \citenamefont {Bobrow}, \citenamefont {Pye}, \citenamefont {Purushotham},\
  and\ \citenamefont {Arnone}}]{Fitzgerald2006}%
  \BibitemOpen
  \bibfield  {author} {\bibinfo {author} {\bibfnamefont {Anthony~J.}\
  \bibnamefont {Fitzgerald}}, \bibinfo {author} {\bibfnamefont {Vincent~P.}\
  \bibnamefont {Wallace}}, \bibinfo {author} {\bibfnamefont {Mercedes}\
  \bibnamefont {Jimenez-Linan}}, \bibinfo {author} {\bibfnamefont {Lynda}\
  \bibnamefont {Bobrow}}, \bibinfo {author} {\bibfnamefont {Richard~J.}\
  \bibnamefont {Pye}}, \bibinfo {author} {\bibfnamefont {Anand~D.}\
  \bibnamefont {Purushotham}}, \ and\ \bibinfo {author} {\bibfnamefont
  {Donald~D.}\ \bibnamefont {Arnone}},\ }\bibfield  {title} {\enquote {\bibinfo
  {title} {Terahertz pulsed imaging of human breast tumors},}\ }\href {\doibase
  10.1148/radiol.2392041315} {\bibfield  {journal} {\bibinfo  {journal}
  {Radiology}\ }\textbf {\bibinfo {volume} {239}},\ \bibinfo {pages} {533--540}
  (\bibinfo {year} {2006})}\BibitemShut {NoStop}%
\bibitem [{\citenamefont {Appleby}\ and\ \citenamefont
  {Wallace}(2007)}]{Appleby2007}%
  \BibitemOpen
  \bibfield  {author} {\bibinfo {author} {\bibfnamefont {R.}~\bibnamefont
  {Appleby}}\ and\ \bibinfo {author} {\bibfnamefont {H.~B.}\ \bibnamefont
  {Wallace}},\ }\bibfield  {title} {\enquote {\bibinfo {title} {Standoff
  detection of weapons and contraband in the 100 {GH}z to 1 {TH}z region},}\
  }\href {\doibase 10.1109/TAP.2007.908543} {\bibfield  {journal} {\bibinfo
  {journal} {IEEE Transactions on Antennas and Propagation}\ }\textbf {\bibinfo
  {volume} {55}},\ \bibinfo {pages} {2944--2956} (\bibinfo {year}
  {2007})}\BibitemShut {NoStop}%
\bibitem [{\citenamefont {Ferguson}\ and\ \citenamefont
  {Zhang}(2002)}]{Ferguson2002}%
  \BibitemOpen
  \bibfield  {author} {\bibinfo {author} {\bibfnamefont {Bradley}\ \bibnamefont
  {Ferguson}}\ and\ \bibinfo {author} {\bibfnamefont {Xi-Cheng}\ \bibnamefont
  {Zhang}},\ }\bibfield  {title} {\enquote {\bibinfo {title} {Materials for
  terahertz science and technology},}\ }\href {\doibase 10.1038/nmat708}
  {\bibfield  {journal} {\bibinfo  {journal} {Nat Mater}\ }\textbf {\bibinfo
  {volume} {1}},\ \bibinfo {pages} {26--33} (\bibinfo {year}
  {2002})}\BibitemShut {NoStop}%
\bibitem [{\citenamefont {Zgirski}\ \emph {et~al.}(2011)\citenamefont
  {Zgirski}, \citenamefont {Bretheau}, \citenamefont {Le~Masne}, \citenamefont
  {Pothier}, \citenamefont {Esteve},\ and\ \citenamefont
  {Urbina}}]{Zgirski2011}%
  \BibitemOpen
  \bibfield  {author} {\bibinfo {author} {\bibfnamefont {M.}~\bibnamefont
  {Zgirski}}, \bibinfo {author} {\bibfnamefont {L.}~\bibnamefont {Bretheau}},
  \bibinfo {author} {\bibfnamefont {Q.}~\bibnamefont {Le~Masne}}, \bibinfo
  {author} {\bibfnamefont {H.}~\bibnamefont {Pothier}}, \bibinfo {author}
  {\bibfnamefont {D.}~\bibnamefont {Esteve}}, \ and\ \bibinfo {author}
  {\bibfnamefont {C.}~\bibnamefont {Urbina}},\ }\bibfield  {title} {\enquote
  {\bibinfo {title} {Evidence for long-lived quasiparticles trapped in
  superconducting point contacts},}\ }\href {\doibase
  10.1103/PhysRevLett.106.257003} {\bibfield  {journal} {\bibinfo  {journal}
  {Phys. Rev. Lett.}\ }\textbf {\bibinfo {volume} {106}},\ \bibinfo {pages}
  {257003} (\bibinfo {year} {2011})}\BibitemShut {NoStop}%
\bibitem [{\citenamefont {Della~Rocca}\ \emph {et~al.}(2007)\citenamefont
  {Della~Rocca}, \citenamefont {Chauvin}, \citenamefont {Huard}, \citenamefont
  {Pothier}, \citenamefont {Esteve},\ and\ \citenamefont
  {Urbina}}]{DellaRocca2007}%
  \BibitemOpen
  \bibfield  {author} {\bibinfo {author} {\bibfnamefont {M.~L.}\ \bibnamefont
  {Della~Rocca}}, \bibinfo {author} {\bibfnamefont {M.}~\bibnamefont
  {Chauvin}}, \bibinfo {author} {\bibfnamefont {B.}~\bibnamefont {Huard}},
  \bibinfo {author} {\bibfnamefont {H.}~\bibnamefont {Pothier}}, \bibinfo
  {author} {\bibfnamefont {D.}~\bibnamefont {Esteve}}, \ and\ \bibinfo {author}
  {\bibfnamefont {C.}~\bibnamefont {Urbina}},\ }\bibfield  {title} {\enquote
  {\bibinfo {title} {Measurement of the current-phase relation of
  superconducting atomic contacts},}\ }\href {\doibase
  10.1103/PhysRevLett.99.127005} {\bibfield  {journal} {\bibinfo  {journal}
  {Phys. Rev. Lett.}\ }\textbf {\bibinfo {volume} {99}},\ \bibinfo {pages}
  {127005} (\bibinfo {year} {2007})}\BibitemShut {NoStop}%
\bibitem [{\citenamefont {Foltyn}\ and\ \citenamefont
  {Zgirski}(2015)}]{Zgirski2015}%
  \BibitemOpen
  \bibfield  {author} {\bibinfo {author} {\bibfnamefont {Marek}\ \bibnamefont
  {Foltyn}}\ and\ \bibinfo {author} {\bibfnamefont {Maciej}\ \bibnamefont
  {Zgirski}},\ }\bibfield  {title} {\enquote {\bibinfo {title} {Gambling with
  superconducting fluctuations},}\ }\href {\doibase
  10.1103/PhysRevApplied.4.024002} {\bibfield  {journal} {\bibinfo  {journal}
  {Phys. Rev. Applied}\ }\textbf {\bibinfo {volume} {4}},\ \bibinfo {pages}
  {024002} (\bibinfo {year} {2015})}\BibitemShut {NoStop}%
\bibitem [{\citenamefont {Chiorescu}\ \emph {et~al.}(2003)\citenamefont
  {Chiorescu}, \citenamefont {Nakamura}, \citenamefont {Harmans},\ and\
  \citenamefont {Mooij}}]{Chiorescu2003}%
  \BibitemOpen
  \bibfield  {author} {\bibinfo {author} {\bibfnamefont {I.}~\bibnamefont
  {Chiorescu}}, \bibinfo {author} {\bibfnamefont {Y.}~\bibnamefont {Nakamura}},
  \bibinfo {author} {\bibfnamefont {C.~J. P.~M.}\ \bibnamefont {Harmans}}, \
  and\ \bibinfo {author} {\bibfnamefont {J.~E.}\ \bibnamefont {Mooij}},\
  }\bibfield  {title} {\enquote {\bibinfo {title} {Coherent quantum dynamics of
  a superconducting flux qubit},}\ }\href {\doibase 10.1126/science.1081045}
  {\bibfield  {journal} {\bibinfo  {journal} {Science}\ }\textbf {\bibinfo
  {volume} {299}},\ \bibinfo {pages} {1869--1871} (\bibinfo {year}
  {2003})}\BibitemShut {NoStop}%
\bibitem [{\citenamefont {Martinis}\ \emph {et~al.}(1987)\citenamefont
  {Martinis}, \citenamefont {Devoret},\ and\ \citenamefont
  {Clarke}}]{Martinis1987}%
  \BibitemOpen
  \bibfield  {author} {\bibinfo {author} {\bibfnamefont {John~M.}\ \bibnamefont
  {Martinis}}, \bibinfo {author} {\bibfnamefont {Michel~H.}\ \bibnamefont
  {Devoret}}, \ and\ \bibinfo {author} {\bibfnamefont {John}\ \bibnamefont
  {Clarke}},\ }\bibfield  {title} {\enquote {\bibinfo {title} {Experimental
  tests for the quantum behavior of a macroscopic degree of freedom: The phase
  difference across a {J}osephson junction},}\ }\href {\doibase
  10.1103/PhysRevB.35.4682} {\bibfield  {journal} {\bibinfo  {journal} {Phys.
  Rev. B}\ }\textbf {\bibinfo {volume} {35}},\ \bibinfo {pages} {4682--4698}
  (\bibinfo {year} {1987})}\BibitemShut {NoStop}%
\bibitem [{\citenamefont {Devoret}\ \emph {et~al.}(2003)\citenamefont
  {Devoret}, \citenamefont {Esteve}, \citenamefont {Urbina}, \citenamefont
  {Martinis}, \citenamefont {Cleland},\ and\ \citenamefont
  {Clark}}]{Devoret2003}%
  \BibitemOpen
  \bibfield  {author} {\bibinfo {author} {\bibfnamefont {M.}~\bibnamefont
  {Devoret}}, \bibinfo {author} {\bibfnamefont {D.}~\bibnamefont {Esteve}},
  \bibinfo {author} {\bibfnamefont {C.}~\bibnamefont {Urbina}}, \bibinfo
  {author} {\bibfnamefont {J.}~\bibnamefont {Martinis}}, \bibinfo {author}
  {\bibfnamefont {A.}~\bibnamefont {Cleland}}, \ and\ \bibinfo {author}
  {\bibfnamefont {J.}~\bibnamefont {Clark}},\ }\bibfield  {title} {\enquote
  {\bibinfo {title} {Macroscopic quantum mechanics of the current-biased
  {J}osephson junction},}\ }in\ \href@noop {} {\emph {\bibinfo {booktitle}
  {Exploring the Quantum Classical Frontier: Recent Advances in Macroscopic
  Quantum Phenomena}}},\ \bibinfo {editor} {edited by\ \bibinfo {editor}
  {\bibfnamefont {J.}~\bibnamefont {Friedman}}\ and\ \bibinfo {editor}
  {\bibfnamefont {S.}~\bibnamefont {Han}}}\ (\bibinfo  {publisher} {Nova
  Science Publishers},\ \bibinfo {year} {2003})\BibitemShut {NoStop}%
\bibitem [{\citenamefont {Wernsdorfer}(2009)}]{Wernsdorfer2009}%
  \BibitemOpen
  \bibfield  {author} {\bibinfo {author} {\bibfnamefont {W}~\bibnamefont
  {Wernsdorfer}},\ }\bibfield  {title} {\enquote {\bibinfo {title} {From
  micro- to nano-squids: applications to nanomagnetism},}\ }\href
  {http://stacks.iop.org/0953-2048/22/i=6/a=064013} {\bibfield  {journal}
  {\bibinfo  {journal} {Superconductor Science and Technology}\ }\textbf
  {\bibinfo {volume} {22}},\ \bibinfo {pages} {064013} (\bibinfo {year}
  {2009})}\BibitemShut {NoStop}%
\bibitem [{\citenamefont {Dubos}\ \emph {et~al.}(2001)\citenamefont {Dubos},
  \citenamefont {Courtois}, \citenamefont {Pannetier}, \citenamefont {Wilhelm},
  \citenamefont {Zaikin},\ and\ \citenamefont {Sch\"on}}]{Dubos2001}%
  \BibitemOpen
  \bibfield  {author} {\bibinfo {author} {\bibfnamefont {P.}~\bibnamefont
  {Dubos}}, \bibinfo {author} {\bibfnamefont {H.}~\bibnamefont {Courtois}},
  \bibinfo {author} {\bibfnamefont {B.}~\bibnamefont {Pannetier}}, \bibinfo
  {author} {\bibfnamefont {F.~K.}\ \bibnamefont {Wilhelm}}, \bibinfo {author}
  {\bibfnamefont {A.~D.}\ \bibnamefont {Zaikin}}, \ and\ \bibinfo {author}
  {\bibfnamefont {G.}~\bibnamefont {Sch\"on}},\ }\bibfield  {title} {\enquote
  {\bibinfo {title} {Josephson critical current in a long mesoscopic {S-N-S}
  junction},}\ }\href {\doibase 10.1103/PhysRevB.63.064502} {\bibfield
  {journal} {\bibinfo  {journal} {Phys. Rev. B}\ }\textbf {\bibinfo {volume}
  {63}},\ \bibinfo {pages} {064502} (\bibinfo {year} {2001})}\BibitemShut
  {NoStop}%
\bibitem [{\citenamefont {Tinkham}(2004)}]{Tinkham2004}%
  \BibitemOpen
  \bibfield  {author} {\bibinfo {author} {\bibfnamefont {M.}~\bibnamefont
  {Tinkham}},\ }\href {https://books.google.pl/books?id=k6AO9nRYbioC} {\emph
  {\bibinfo {title} {Introduction to Superconductivity: Second Edition}}},\
  Dover Books on Physics\ (\bibinfo  {publisher} {Dover Publications},\
  \bibinfo {year} {2004})\BibitemShut {NoStop}%
\bibitem [{\citenamefont {Courtois}\ \emph {et~al.}(2008)\citenamefont
  {Courtois}, \citenamefont {Meschke}, \citenamefont {Peltonen},\ and\
  \citenamefont {Pekola}}]{Courtois2008}%
  \BibitemOpen
  \bibfield  {author} {\bibinfo {author} {\bibfnamefont {H.}~\bibnamefont
  {Courtois}}, \bibinfo {author} {\bibfnamefont {M.}~\bibnamefont {Meschke}},
  \bibinfo {author} {\bibfnamefont {J.~T.}\ \bibnamefont {Peltonen}}, \ and\
  \bibinfo {author} {\bibfnamefont {J.~P.}\ \bibnamefont {Pekola}},\ }\bibfield
   {title} {\enquote {\bibinfo {title} {Origin of hysteresis in a proximity
  {J}osephson junction},}\ }\href {\doibase 10.1103/PhysRevLett.101.067002}
  {\bibfield  {journal} {\bibinfo  {journal} {Phys. Rev. Lett.}\ }\textbf
  {\bibinfo {volume} {101}},\ \bibinfo {pages} {067002} (\bibinfo {year}
  {2008})}\BibitemShut {NoStop}%
\bibitem [{\citenamefont {Maisi}\ \emph {et~al.}(2013)\citenamefont {Maisi},
  \citenamefont {Lotkhov}, \citenamefont {Kemppinen}, \citenamefont {Heimes},
  \citenamefont {Muhonen},\ and\ \citenamefont {Pekola}}]{Maisi2013}%
  \BibitemOpen
  \bibfield  {author} {\bibinfo {author} {\bibfnamefont {V.~F.}\ \bibnamefont
  {Maisi}}, \bibinfo {author} {\bibfnamefont {S.~V.}\ \bibnamefont {Lotkhov}},
  \bibinfo {author} {\bibfnamefont {A.}~\bibnamefont {Kemppinen}}, \bibinfo
  {author} {\bibfnamefont {A.}~\bibnamefont {Heimes}}, \bibinfo {author}
  {\bibfnamefont {J.~T.}\ \bibnamefont {Muhonen}}, \ and\ \bibinfo {author}
  {\bibfnamefont {J.~P.}\ \bibnamefont {Pekola}},\ }\bibfield  {title}
  {\enquote {\bibinfo {title} {Excitation of single quasiparticles in a small
  superconducting al island connected to normal-metal leads by tunnel
  junctions},}\ }\href {\doibase 10.1103/PhysRevLett.111.147001} {\bibfield
  {journal} {\bibinfo  {journal} {Phys. Rev. Lett.}\ }\textbf {\bibinfo
  {volume} {111}},\ \bibinfo {pages} {147001} (\bibinfo {year}
  {2013})}\BibitemShut {NoStop}%
\bibitem [{\citenamefont {Phillips}(1959)}]{Phillips1959}%
  \BibitemOpen
  \bibfield  {author} {\bibinfo {author} {\bibfnamefont {Norman~E.}\
  \bibnamefont {Phillips}},\ }\bibfield  {title} {\enquote {\bibinfo {title}
  {Heat capacity of aluminum between 0.1{K} and 4.0{K}},}\ }\href {\doibase
  10.1103/PhysRev.114.676} {\bibfield  {journal} {\bibinfo  {journal} {Phys.
  Rev.}\ }\textbf {\bibinfo {volume} {114}},\ \bibinfo {pages} {676--685}
  (\bibinfo {year} {1959})}\BibitemShut {NoStop}%
\bibitem [{not()}]{note}%
  \BibitemOpen
  \href@noop {} {}\bibinfo {note} {${\Delta} {T}={Q}/{C}$, temperature rise of
  the absorber ${\Delta} {T}$ upon absorption of a small energy ${Q}$ is
  inversely proportional to the heat capacity of the absorber ${C}$ ({HIGH
  SENSITIVITY}); $\tau ={C}/{G}$, relaxation time of the absorber is
  proportional to its heat capacity ${C}$ with ${G}$ being heat conductance
  ({FAST RESPONSE})}\BibitemShut {NoStop}%
\end{thebibliography}%

\section*{Acknowledgments}
The work is supported by Foundation for Polish Science (First TEAM/2016-1/10), the EAgLE project (FP7-REGPOT-2013-1, Project No. 316014) and Polish Ministry of Science and Higher Education (Grant 2819/7.PR/2013/2). We acknowledge the availability of the facilities and technical support provided by Otaniemi research infrastructure for Micro and Nanotechnologies (OtaNano).

\section*{Additional information}
The authors declare no competing financial interests. Correspondence and requests for materials should be addressed to M.Z.

\end{document}